# LabelImg: CNN-Based Surface Defect Detection

Mohsen Asghari Ilani[1], Yaser Mike Banad[1]

[1] School of Electrical and Computer Engineering, University of Oklahoma, Norman, 73019, U.S.A.

**Abstract**

In the journey of computer vision system development, the acquisition and utilization of annotated images play a central role, providing information about object identity, spatial extent, and viewpoint in depicted scenes. However, thermal manufacturing processes like Laser Powder Bed Fusion (LPBF) often yield surfaces with defects such as Spatter, Crack, Pinhole, and Hole due to the Balling phenomenon. Preprocessing images from LPBF, riddled with defects, presents challenges in training machine learning (ML) algorithms. Detecting defects is critical for predicting production quality and identifying crucial points in artificial or natural structures. This paper introduces a deep learning-based approach utilizing Convolutional Neural Networks (CNNs) to automatically detect and segment surface defects like cracks, spatter, holes, and pinholes on production surfaces. In contrast to traditional machine learning techniques requiring extensive processing time and manual feature crafting, deep learning proves more accurate. The proposed architecture undergoes training and testing on 14,982 labeled images annotated using the LabelImg tool. Each object in the images is manually annotated with bounding boxes and segmented masks. The trained CNN, coupled with OpenCV preprocessing techniques, achieves an impressive 99.54% accuracy on the dataset with resolutions of 1536 × 1103 pixels. Evaluation metrics for 50 true crack tests demonstrate precision, recall, and F1-score exceeding 96%, 98%, and 97%, respectively. Similarly, for 124 true pinhole tests, the metrics are 99%, 100%, and 100%, for 258 true hole tests, they are 99%, 99%, and 99%, and for 318 spatter tests, the metrics are 100%, 99%, and 100%. These results highlight the precision and effectiveness of the entire process, showcasing its potential for reliable defect detection in production surfaces.

## Introduction

Monitoring and maintaining structural production fabricated by different manufacturing process, such as surface defect detection like crack, hole, pinhole and spatter, is essential in recognizing anomalies as a clear signal for potential defects. Several artificial structures can significantly benefit from the early sensing of possible physical defects in them via automatic object detection. Crack detection along with the other possible defects are already in practice for a long time but due to technical advancements and innovative monitoring, the transformation of manual procedures into automatic ones is essential. The automation would reduce inspection costs and human errors while promoting inspection speed while increasing the reliability for real-time applications. Various image-processing techniques (IPTs) had been used in the past to address the problem [1–5]. Such techniques were helpful in feature identification but suffered while distinguishing samples with genuine cracks with lighting effects, edges, and shadows and other nondestructive methodologies [6]. It is almost true for most techniques use the ultrasonic, eddy current and other strategies used to detect the seen and unseen defects. Furthermore, such techniques are inflexible to various image types and formats. Since object detection in surfaces are visual anomalies, researchers and inspectors have identified them as computer vision problems that can be solved using deep learning techniques.

Deep learning is inspired by its escalating successful implementation and state-of-the-art solutions existing across multiple domains [7-12]. Computer vision through neural networks and their specially designed architectures have successfully replaced the classical, manually handcrafted methods and approaches [13-17]. Convolutional Neural networks (CNNs) are specially designed neural networks to handle and work with image data (spatial 2-Dimensional data). CNN is mighty in extracting spatial-visual features from images (from lower-level features to higher-level features), resulting in increased performance speed and accuracy for computer vision tasks. Therefore, they excel in various computer vision applications like image classification, image segmentation, object detection, etc. CNN can be applied to perform automatic surface defect detection like crack, spatter, hole, and pinhole on the surface image dataset by combining various convolution and pooling layers [18–20].

In the practical investigations, the availability of visual data has experienced a dramatic change in the last decade, especially via the Internet, which has given researchers access to billions of images and videos. While large volumes of pictures are available or extracted from an experimental work, building a large data set of annotated images with many objects still constitutes a costly and lengthy endeavor. Traditionally, data sets are built by individual research groups and are tailored to solve specific problems. Therefore, many currently available data sets used in computer vision only contain a small number of object classes, and reliable detectors exist for a few of them. Notable recent exceptions are the Caltech 101 data set, with 101 object classes (later extended to 256 object classes), ImageNet, the PASCAL collection containing 20 object classes, the CBCL-street scenes database, comprising eight object categories in street scenes, and the database of scenes from the Lotus Hill Research Institute.

This challenge is extended when the surface defects fabricated by thermal manufacturing processes including of instability in cooling, grain growth, phase changes from FCC/BCC to HCP leading to apply strength inhomogenously in material crystal, along with laser beam diffraction owing to spatter melted materials, all have made the Laser Powder Bed Fusion (LPBF) as a manufacturing process with a pile of surface defects. In the practical investigations, the fabrication of alloys in LPBF encounters great difficulties, including the incurrence of pores, interface cracks between reinforced particles and metal matrix [21], and thermal residual stress [22]. The reinforcements have a negative effect on the densification behavior in LPBF. Balling phenomenon is considered as one of the causes of hole and pinhole in the LPBF of MMCs due to the increased viscosity by the reinforcements [23]. For MMCs with a higher volume fraction of the reinforcements, the balling phenomenon would be more prominent. In the fabrication of GWs with 25 vol% diamond grains, severe balling materials were observed on the top surface, resulting in many large holes [24]. Even though pore is an essential element in GWs, large pores are undesirable because of the reduced mechanical strength of fabricated material. It was reported that an increased laser energy density allows fabricating denser parts. Nevertheless, an excessive laser energy input would cause higher thermal residual stress and cracks between the reinforced particles and the metal matrix due to the distinct thermal expansion coefficients. The balling phenomenon caused by the reinforcements may also be associated with particle spattering. In the fabrication of MMCs, the reinforcements may turn into spatters under the impact of high-speed metal vapor in LPBF, especially for the infusible reinforcements. Li et al. [21] observed the diamond spattering in the fabrication of GWs in LPBF and revealed the detrimental effect on the formation quality. Coarser reinforcements may induce more prominent spattering phenomenon. Due to the larger diamond grains and higher volume percentage, spattering plays a critical role in the balling phenomenon in the LPBF of GWs. However, there are few studies on the spattering behavior of reinforcements and its effect on the formation mechanisms of MMCs in LPBF [25].

This article delves into instance segmentation, a novel approach generating segmentation bounding boxes around objects in input images through image preprocessing in OpenCV and LabelImg. A unique aspect is the focus on instance segmentation in a four-class object detection dataset related to surfaces produced by electro-thermal manufacturing processes. Unlike previous works concentrating on drawing bounding boxes around surface cracks, our approach manually labels each object using bounding boxes and detailed shape-like rectangles in LabelImg. This meticulous labeling enhances the visualization of essential object regions, improving output generation.

The dataset, comprising 14,982 images, is collected and labeled for instance segmentation. Our proposed architecture automates the identification of surface cracks, spatter, holes, and pinholes. Contrary to algorithms proposed years ago that struggle with noisy data and require manual parameter adjustment, our approach utilizes deep learning and optimization techniques for efficient and successful four-class detection (crack, spatter, pinhole, and hole). This strategy eliminates the need for manual parameter adjustments, addressing limitations in traditional algorithms associated with varying appearances and multiple clusters in real-world images of alloys. The result is a more robust and automatic analysis of large datasets.

The objective of this study is to employ LabelImg for annotating objects on a surface that exhibits numerous defects. LabelImg is selected as the annotation tool due to its lightweight nature and user-friendly interface, facilitating the labeling of object bounding boxes in images. This article offers an overview of LabelImg, discussing its utility, and provides guidance on annotating images with ease. The choice of appropriate image annotation software is crucial for the sustained success of computer vision applications. Consequently, this guide is designed to assist in the evaluation process to identify the most suitable image annotation software for both current and future projects.

The study specifically focuses on the development of a CNN classifier aimed at categorizing additive manufacturing (AM) components based on their surface images. This approach presents a more straightforward and convenient means for automatically identifying surface conditions, subsequently informing decisions on polishing conditions. While existing polishing methods have demonstrated effectiveness in enhancing the surface quality of metal AM components, they possess inherent limitations, primarily related to challenges in controlling the polishing process. In response to these challenges, LabelImg is employed in this study to label images based on their geometrical shape, allowing for the optimal selection of bounding boxes and annotation labels. This choice is motivated by the complex shapes and varying quantities of defects on the surface, each exhibiting different types.

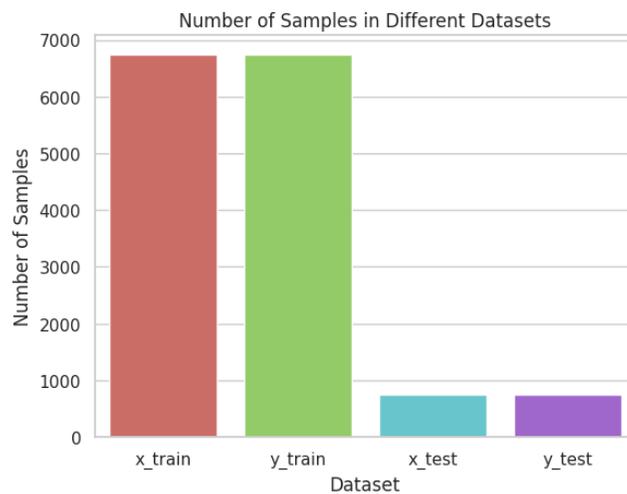

**Figure 1.** Training and Test Datasets.

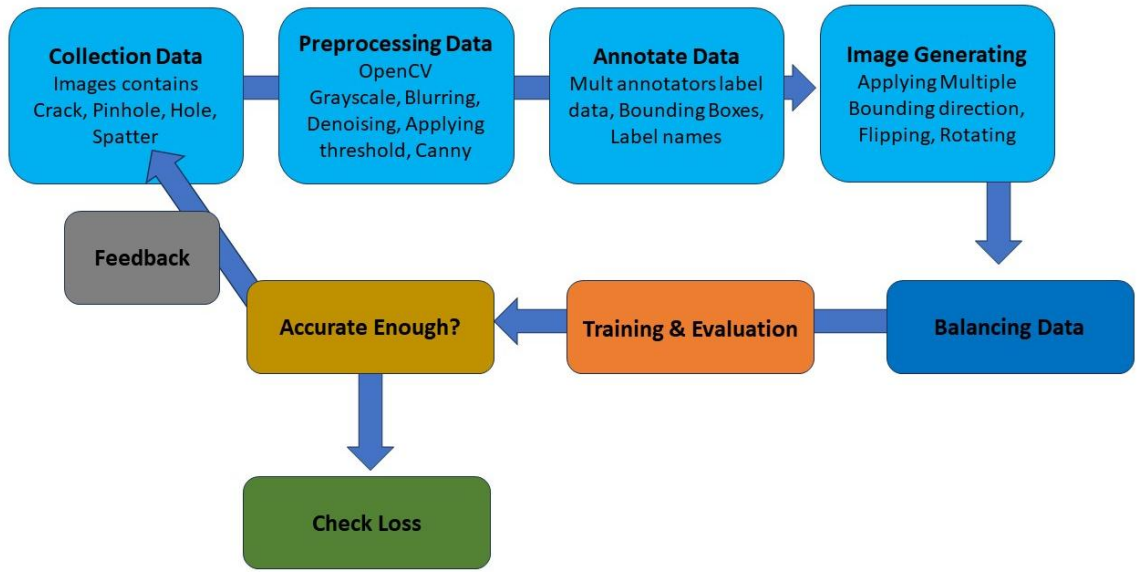

**Figure 2.** LabelImg Approach Strategy on CNN-based Model.

# Materials and Methods

Being a nascent domain in machine learning, deep learning exhibits considerable promise in defect detection. This is achieved through the continual reduction in dimensions during feature learning, mitigating the impact of feature extraction on identification outcomes and thereby enhancing the accuracy of defect detection. The integration of deep learning models with digital image processing techniques is a prevalent approach across diverse fields, validated by their remarkable performance in tasks such as environment recognition and autonomous vehicle control [26]. Consequently, these models can be effectively applied to detect anomalies in welding processes within additive manufacturing (AM). Xiong and Ding employed a deep neural network for predicting bead geometry in relation to various process variables [16, 17]. However, their focus on dimensional values such as width, height, and toe angle limited the prediction to geometrical aspects, excluding potential variations indicative of bead defects. To broaden the scope of anomaly detection, researchers have increasingly turned to utilizing image data from various vision sensors [27–32].

The study delved into the performance and generalizability of Convolutional Neural Networks (CNNs), capitalizing on their ability to preserve spatial relationships between voxels in extracted 3D image tiles. The learnable parameters of a CNN scale with filter size and the number of filters, rather than the number of input features. Consequently, augmenting the number of layers does not significantly increase the number of learnable parameters [33]. Following experimentation with different configurations such as varying layer counts, kernel sizes, and pooling options, the CNN architecture illustrated in Fig. 8, akin to our approach for detecting surface defects, was chosen for all subsequent training and testing. This selection was based on demonstrating the highest average classification accuracy. The CNN architecture, post the layerwise image tile input layer, comprises three sets of 3D Convolution (CONV), Batch Normalization (BN), Rectified Linear Unit (ReLU), and Max Pooling (MP) layers before reaching the Fully Connected (Dense), Softmax, and Classification layers. During each convolutional layer, diverse 3D convolution filters were applied to the tile for feature extraction within the input dataset [36]. The resulting tile was transformed into a dataset with the same 3D size as the input data, and the outcomes of each filter were concatenated along a fourth dimension.

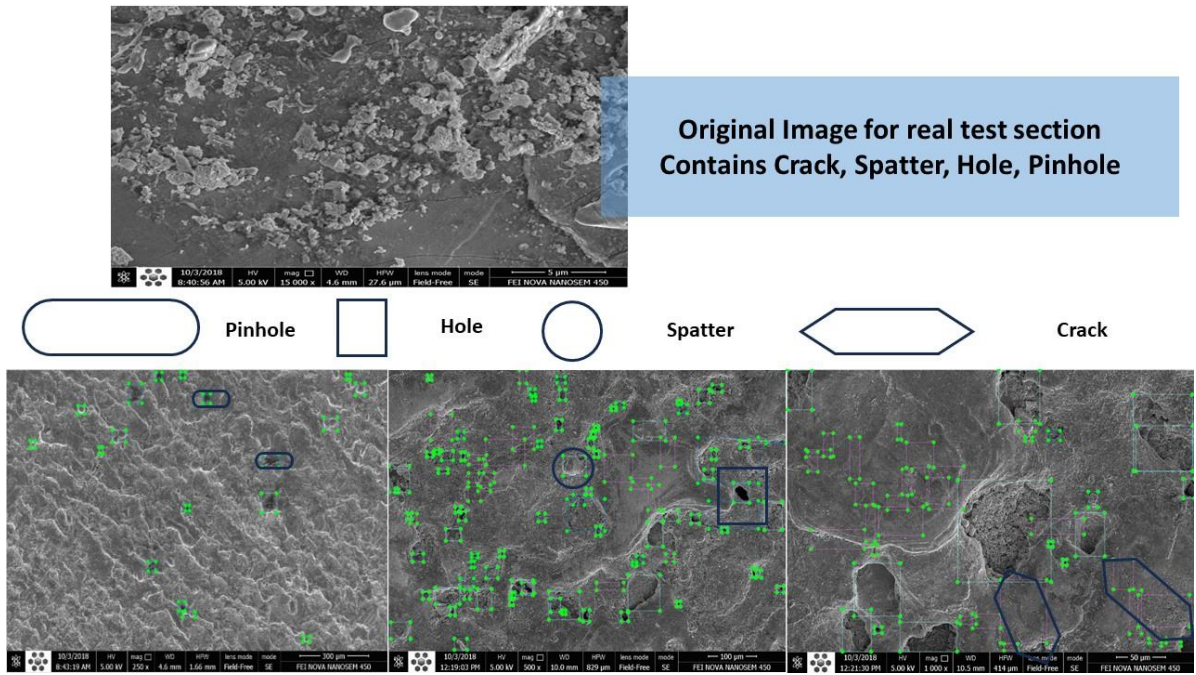

**Figure 3.** LabelImg Annotation Extraction (Bounding boxes + Label)

Subsequently, the results from the convolutional layers underwent batch normalization to maintain a mean of zero and unit variance. This was followed by a ReLU layer, converting negative values to zero and leaving nonnegative values unchanged [33]. Finally, the ReLU layer's output fed into the max-pooling layer. The max-pooling layer employed a 2 × 2 × 2 filter, sliding across the input dataset. The largest value within the filter was added to the output dataset, effectively reducing data size while preserving spatial invariance and minimizing data loss. The sequence of CONV-BN-ReLU-MP was iterated three times, progressively reducing the size of the 3D input tile and using smaller convolutional filters. In the last convolutional group, the max-pooling layer was omitted, and the output from the final ReLU layer was input into a fully connected neural network. The dense layer's result determined the classification probabilities in the softmax layer, leading to the final classification layer.

The annotations generated using the LabelImg tool were saved as XML files. Following this, a series of image preprocessing steps, including resizing, cleaning, refining, and denoising, were applied to the annotated images. These processed datasets are then prepared for the training and testing split phase. In ***Figure 3***, the extraction of bounding boxes and labels from the annotated images is demonstrated, and a comparison with non-labeled images is presented. It is noteworthy that the distribution of labels is not uniform during the annotation process.

To address the imbalance in label distribution while training the Convolutional Neural Network (CNN) model on the training and validation datasets, a weighting strategy is applied, as illustrated in the ***Figure 4***. This involves assigning weights to labels based on their distribution in the datasets. The purpose of this weighting is to ensure that the model gives adequate attention to less frequent classes, thereby improving its ability to generalize across all label categories. The image preprocessing steps, along with the weighting strategy for label distribution, contribute to the optimization of the datasets for training and testing the CNN model. The visualization in Figure 4 provides insight into the extraction of bounding boxes and labels, highlighting the impact of preprocessing and the consideration of label distribution for effective model training.

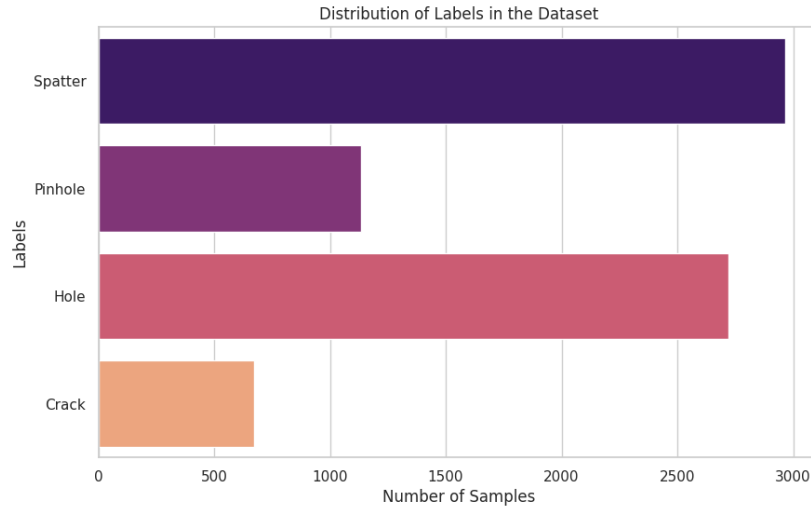

**Figure 4.** Balancing Annotation Labelled by LabelImg.

The model employs a series of Convolutional 2D (Conv2D) layers followed by MaxPooling2D layers, forming a hierarchical feature extraction process. The initial Conv2D layer, with a kernel size of 3x3, processes the input, resulting in an output shape of (120, 80, 64). Subsequent MaxPooling2D, with a 2x2 pool size, reduces the spatial dimensions by half, yielding an output of (60, 40, 64). This pattern repeats with additional Conv2D and MaxPooling2D layers, progressively increasing the depth and complexity of feature representation. The second Conv2D layer amplifies the feature depth to 512, operating on the (60, 40, 64) input shape and leading to an output shape of (60, 40, 512). The associated MaxPooling2D layer further downscales the spatial dimensions, resulting in (20, 14, 512). The subsequent Conv2D layers maintain the same pattern, refining the feature hierarchy, as shown in *Figure 5*.

As the architecture advances, the final Conv2D layer with a kernel size of 3x3 operates on a reduced spatial resolution of (5, 4, 256), followed by MaxPooling2D, yielding a condensed (2, 2, 256) output. A Dropout layer is introduced to mitigate overfitting, contributing to model robustness by randomly deactivating a fraction of neurons during training. The total number of parameters, a crucial metric indicating the model's complexity, amounts to 3,901,892, with 3,901,764 being trainable. This implies the adaptability of the network during the training process. The non-trainable parameters, amounting to 128, likely pertain to biases within the network layers.

## Results and Discussion

this investigation takes advantage of annotated objects provided by the LabelImg tool, enhancing the accuracy of object identification compared to relying solely on the model's training. *Figure 5* succinctly presents an overview of the training and validation outcomes. The dataset maintains a balanced distribution of defects and intact images at a 1:1 ratio, with a training-validation split of 4:1. The training accuracy is computed from a dataset of 6,742 images, while the validation set constitutes 20% of the training data. Notably, exceptional accuracy is attained, peaking at 99.54% during the 51st epoch for training and 97.95% at the 49th epoch for validation. The utilization of two GPUs significantly accelerates the training process, reducing the time required to reach the 100th epoch by approximately 90 minutes. It is essential to highlight that the estimated running time on a CPU alone exceeds 6 hours, emphasizing the computational efficiency gained through GPU acceleration. Conversely, the training loss, illustrated in *Figure 6*, serves as an evaluation of the model's fidelity to the training data, indicating its proficiency in capturing patterns within the dataset. In contrast, the validation loss assesses the model's ability to generalize to new, unseen data.

In our study, the training process of the Convolutional Neural Network (CNN) over 100 epochs, the model demonstrated a notable evolution in its performance, reflected in the decreasing loss and increasing accuracy for both training and validation datasets. Initially, during the first epoch, the loss on the training data was 0.5820, with an accuracy of 25.53%, while on the validation set, the loss was 1.2901 with an accuracy of 40.40%. The subsequent epochs showed a consistent trend of diminishing loss values and ascending accuracy scores. The final epoch, epoch 100, showcased remarkable convergence, with minimal loss values of 0.0030 on the training set and 0.0166 on the validation set, accompanied by high accuracies of 99.65% and 99.26%, respectively. This convergence indicates the model's successful learning and generalization capabilities over the training period.

The CNN model demonstrated highly efficient performance in detecting various surface defects, as evidenced by precision, recall, and F1-score metrics shown in *Table 1*. Specifically, crack detection achieved a precision of 0.96, recall of 0.98, and an F1-score of 0.97, indicating a robust identification of this defect. Pinhole detection exhibited exemplary results with a precision of 0.99, recall of 1.00, and a perfect F1-score of 1.00, showcasing the model's accuracy in identifying pinhole defects. Similarly, hole detection achieved a precision of 0.99, recall of 0.99, and an F1-score of 0.99, emphasizing the model's efficiency in recognizing hole defects. The Spatter category showed outstanding results, with a precision of 1.00, recall of 0.99, and a perfect F1-score of 1.00, underscoring the model's exceptional performance in detecting spatter defects.

Moreover, the accuracy of the model across all defect classes reached 0.99, providing a comprehensive measure of its effectiveness in identifying defects within the dataset. The micro-averaged and weighted-average metrics further support the model's consistent and high-level performance across the various defect categories. The reported values of 0.99 for precision, recall, and F1-score for both micro and weighted averages reinforce the model's proficiency in efficiently detecting and classifying crack, pinhole, hole, and spatter defects within the given dataset comprising 750 instances.

The overall prediction accuracy of the model is presented using a confusion matrix, which summarizes the number of accurate predictions among all class labels in a matrix form. Here, the correlation between predicted labels (x-axis) and true labels (y-axis) is presented. The diagonal blue (from light blue to dark blue) elements show the number of right predictions out of 50 data for crack detection, 124 data for crack detection, 258 data for crack detection, and 318 data for crack detection. The white elements show the inaccurate predictions corresponding to each class. For instance, when the surface was level 1, it was predicted correctly in 49 out of 50 instances and it is inaccurately predicted as level 2, 0 out of 124, for level 2, 3 out of 258 and for level 3, 2 out of 318. Overall, the results are presented the more than %99 accuracy in surface defect prediction.

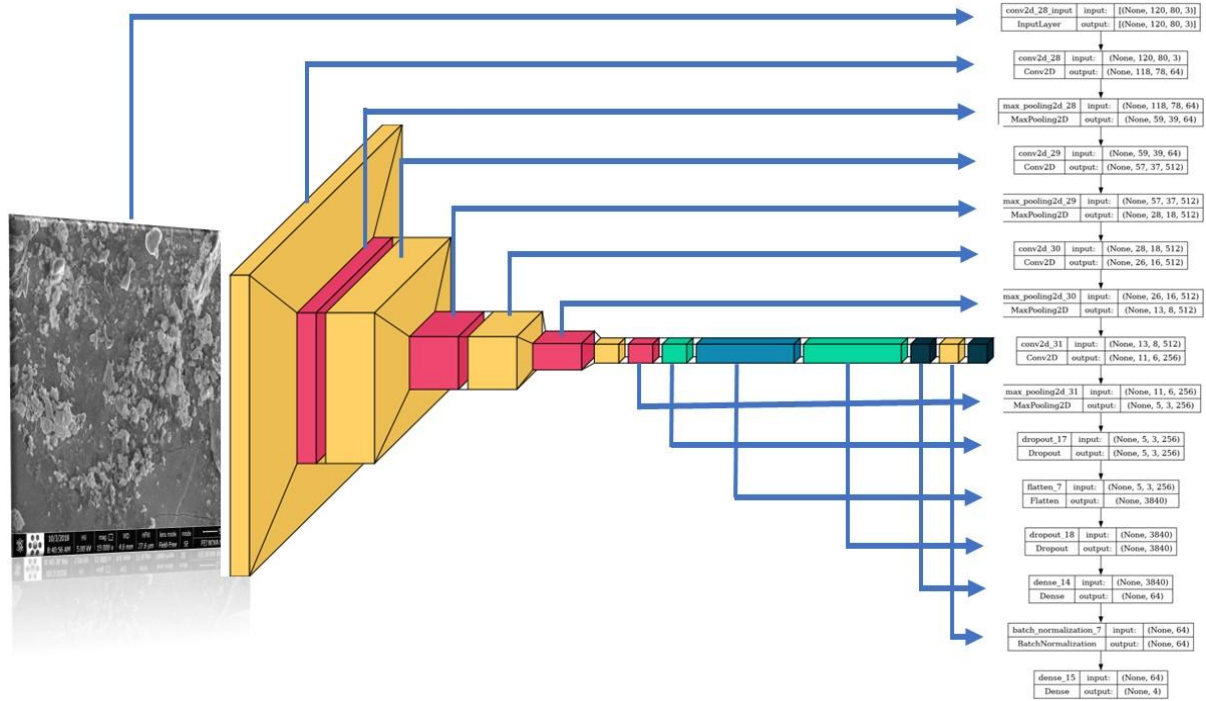

**Figure 5.** Schematic of CNN-based Architecture.

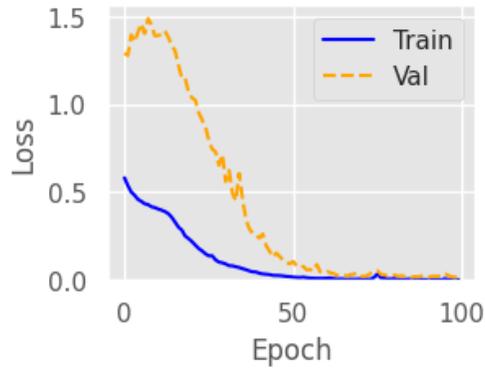

**Figure 6.** Training and Validation Loss per Epoch.

**Table 1.** Evaluation of CNN-based Model.

| Label | Number | precision | Recall | F1-score | Support |
|---|---|---|---|---|---|
| **Crack** | 0 | 0.96 | 0.98 | 0.97 | 50 |
| **Pinhole** | 1 | 0.99 | 1.00 | 1.00 | 124 |
| **Hole** | 2 | 0.99 | 0.99 | 0.99 | 258 |
| **Spatter** | 3 | 1.00 | 0.99 | 1.00 | 318 |
| **Accuracy** | | | | 0.99 | 750 |
| **Micro Ave** | 0.99 | 0.99 | 0.99 | 0.99 | 750 |
| **Weighted Ave** | 0.99 | 0.99 | 0.99 | 0.99 | 750 |

# Conclusion

The study introduces a novel LabelImg approach to enhance the accuracy of surface defect detection in additively manufactured (AM) components using CNN-based intelligence. The presence of defects such as cracks, spatter, holes, and pinholes, attributed to the Balling phenomena, prompts the utilization of a CNN-based approach. Following the labeling of defects, the annotation (bounding boxes and labels) is separated for feeding into the training and testing phases. This annotated data also serves as feedback for refining preprocessing hyperparameters. The results underscore the efficacy of the deep learning approach in accurately categorizing AM-built surfaces based on their morphology, promising substantial applications in industry by saving time and costs associated with manual inspection and quality checks. Key findings from the study include:

- A CNN model is successfully trained to predict the surface category of AM components, achieving a prediction accuracy exceeding 99%.
- The CNN-based model on AM-built images demonstrates complexity in crack types, with precision at 96%, recall at 98%, and an F1-score of 97%. The study suggests exploring crack type separation for enhanced precision in future investigations.
- Balancing datasets significantly improves training and validation loss and accuracy from 32% to 99%.
- Employing OpenCV for image processing, incorporating techniques such as GaussianBlur, fastNlMeansDenoising, adaptiveThreshold, and Canny, facilitates easier defect identification by the model.
- Evaluation metrics exceeding 99% in precision, recall, and F1-score for four classifications underscore the precision and effectiveness of the entire process, from image preprocessing to training and evaluation.
- LabelImg is identified as a potential tool and strategy for images with intricate details.

While the current approach shows promise, future improvements are suggested. The model's accuracy and robustness depend on the quality of the training data, indicating a need for a more extensive dataset encompassing various materials and wider parameter ranges. Additionally, the model's capabilities can be extended to identify internal defects by incorporating advanced imaging techniques like computed tomography (CT). Finally, there is potential for the method to evolve into a real-time/in situ approach for surface classification by integrating suitable vision sensors within the build chamber.